\documentclass{aa}

\usepackage{graphicx}
\usepackage{natbib}
\bibpunct{(}{)}{;}{a}{}{,}

\newcommand{\Lsun}{L$_{\odot}$}
\newcommand{\Msun}{M$_{\odot}$}
\newcommand{\Rsun}{R$_{\odot}$}

\newcommand{\mic}{$\mu$m}

\newcommand{\halfa}{H$\alpha$}
\def\specchar#1{\uppercase{#1}}
\def\HeI{\mbox{He\,\specchar{i}}} 
\def\cdrev#1{#1}
\def\cdrm#1{}
\begin{document}

\title{The absence of the 10~\mic \ silicate feature in the isolated Herbig Ae star
HD~100453}

   \author{G. Meeus
          \inst{1}
          \and
          J. Bouwman\inst{2}
                    \and
          C. Dominik\inst{3}           
                    \and
          L.B.F.M. Waters\inst{3}
                    \and
          A. de Koter\inst{3}
}

   \offprints{G. Meeus \\ (e-mail: gwendolyn@aip.de)}

   \institute{Astrophysikalisches Institut Potsdam (AIP), An der Sternwarte 16, 
    D-14482 Potsdam\\
         \and
             Service d'Astrophysique, CEA Saclay, F-91191 Gif-sur-Yvette, France\\
         \and
             Astronomical Institute Anton Pannekoek, University of 
Amsterdam, Kruislaan 403, NL-1098 SJ Amsterdam, The Netherlands
                 }

   \date{Received ...; accepted ...}

   \abstract{We analyse the optical and IR spectra, as well as the spectral
energy distribution (UV to mm) of the candidate Herbig Ae star HD~100453. 
This star is particular, as it shows an energy distribution similar to that 
of other isolated Herbig Ae/Be stars (HAEBEs), but unlike most of them, it 
does not 
have a silicate emission feature at 10~\mic, as is shown in \citet{meeus2001}. 
We confirm the HAEBE nature of HD~100453 through an analysis of its optical 
spectrum and derived location in the H-R diagram. The IR spectrum of HD~100453
is modelled by an optically thin radiative transfer code, from which we derive 
constraints on the composition, grain-size and temperature distribution 
of the circumstellar dust. We show that it is both possible to explain the lack 
of the silicate feature as (1) a grain-size effect - lack of {\bf small} silicate 
grains, and (2) a temperature effect - lack of small, {\bf hot} silicates, as 
proposed by \citet{dullemond2001}, and discuss both possibilities. \cdrm{Finally, we 
show that the latter possibility is the more preferable.}
  
\keywords{circumstellar matter - stars: pre-main sequence; individual: 
             HD\,100453 - ISM: lines and bands}
   }

\authorrunning{G. Meeus et al.}
\titlerunning{Absence of the 10~\mic \ silicate feature in HD100453}
   \maketitle


\section{Introduction}

Herbig Ae/Be stars are intermediate-mass pre-main sequence stars which show
an IR excess due to circumstellar (CS) dust. This dust is believed to be
located in a disc \citep[see e.g.][]{adams1987ApJ...312..788A,beckwith1990AJ.....99..924B,meeus1998A&A...329..131M}; 
\citet{chiang1997ApJ...490..368C} were the first to develop a CS disc model in
which a warm layer of optically thin material surrounds the (cold)
midplane. This passive disc model provides an elegant solution to
the puzzling observation that the disc must be optically thick in
the near-IR and mid-IR (in order to account for the sub-millimeter
flux) while the Si-O stretch of amorphous silicates at 9.7 $\mu$m
is observed to be in emission (implying it is caused by optically
thin dust).

Since small amorphous silicate grains are the most abundant dust
species in interstellar space, their spectral signature is
expected to be present during virtually all stages of the star
formation process: in absorption during the protostellar collapse
and active accretion disc phases, and in emission during the
passive disc and debris disc phases. Indeed, observations of star
forming regions confirm this picture. However, during our analysis
of the infrared spectra of passive discs surrounding isolated
Herbig Ae/Be stars taken with the Infrared Space Observatory 
\citep[ISO;][]{kessler1996A&A...315L..27K}, we found that some stars do not 
show any evidence for the presence of the 9.7 $\mu$m band \citep[][ hereafter 
Paper~I]{meeus2001}. In Paper~I we speculate that the lack of
silicate emission is due to the absence of small silicate grains
in the disc, due to grain growth and the removal of small grains
by radiation pressure.

\citet[][ hereafter DDN]{dullemond2001} propose a different
explanation, by avoiding the presence of small silicate grains in
a certain temperature range. In the DDN model, the disc inner rim
has a large scale-height due to the fact that it receives direct
stellar radiation (an effect not taken into account in the Chiang
\& Goldreich models). The inner rim causes a shadowed region
behind the rim where the temperature is low; at some distance from
the star the surface of the flaring disc emerges from the shadow
and receives direct starlight. The rim causes two effects: (1) it
creates a prominent near-IR flux contribution which is in good
agreement with observations \citep[see][ DDN]{natta2001A&A...371..186N}, and (2)
for certain rim heights the shadowed region can suppress the
strength of the 9.7 $\mu$m silicate emission (see DDN).

Stimulated by these considerations, we decided to derive empirical
constraints on the amount of mass that can be present in the form
of warm optically thin silicate grains in HAEBE stars that lack
the silicate feature. We investigate the effect the size and the temperature 
distribution of dust grains can have on the appearance of spectral features. 
Therefore, we look for observational constraints on the average size of the 
silicate grains and the maximal mass of small, warm silicate grains. This we 
do by modelling the isolated Herbig Ae star HD~100453, which is one of the 
four isolated HAEBEs in our sample where the silicate feature is found to be 
absent \citep{meeus2001}. We selected this star because it is the brightest 
of all four, resulting in observations with higher signal to noise ratios. The 
model we used to derive the constraints is the optically thin radiative 
transfer code MODUST (Bouwman \& de Koter, in preparation).


\section{Earlier and new observations}

\begin{table*}
\caption{The stellar parameters of HD~100453, based on optical spectra (obtained 
with Feros), data from the literature and Hipparcos measurements (van den Ancker, 
private communication).}
\label{para}
\begin{center}
\begin{tabular}{cccccccccccc}
\hline
\hline
Spectral&V    &E[B-V]& T$_{\mathrm{eff}}$&log g &L & L$_{\mathrm{IR}}$/L$_{*}$&R &Age&Parallax&Distance &v $\sin$ i \\
Type    &(mag)&& (K)               &      &(\Lsun)&&(\Rsun)&(Myr)&(\arcsec)&(pc)&(kms$^{-1}$)\\[1ex]
\hline
\rule[3mm]{0mm}{1mm}
A9Ve    &7.79 & 0.08&7500              & 4.5   &$10^{+2.1}_{-1.5}$&0.77& 1.86&$\ga$ 10 &$8.76^{+0.76}_{-0.76}$ &$114^{+11}_{-9}$ &$56^{+5}_{-5}$\\[1ex]
\hline
\end{tabular}
\end{center}
\end{table*}

\begin{figure}
\resizebox{\hsize}{!}{{\rotatebox{90}{\includegraphics{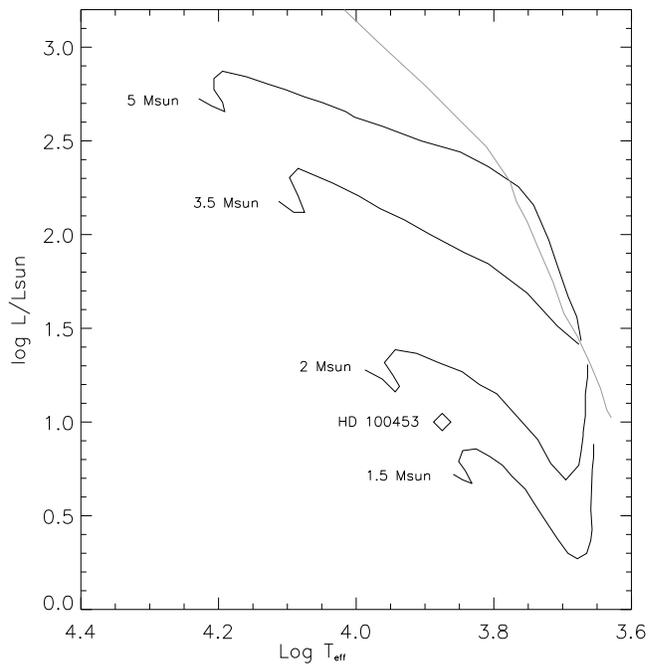}}}}
\caption{PMS evolutionary tracks in the HR diagram, labelled with the respective 
mass (dark lines). Grey line: 
the birthline for an accretion rate of 10$^{-5}$ \Msun yr$^{-1}$. The position of 
the Herbig Ae star HD~100453 is indicated with a diamond. The star is located 
closely to the ZAMS, suggesting an evolved PMS stage. The luminosity
is log L/\Lsun = 1.00$^{+0.08}_{-0.07}$ (van den Ancker, private communication). 
Tracks are from \citet{palla1993ApJ...418..414P}.}
\label{hrd}
\end{figure}

The star HD~100453 was classified as a planetary nebula by
\citet{garcia1997A&AS..126..479G}. However, recent observations show it to be 
an A9Ve type star, belonging to the group of the isolated HAEBEs 
\citep[][]{meeus2001}, which are considered as the more evolved members of the 
group. Hipparcos measurements give a parallax of 8.76 mas, from which a distance 
of 114 pc is derived. We located the star in the Hertzsprung-Russell-diagram (see 
Fig.~\ref{hrd}). Its position is close to the ZAMS; its age is estimated to be 
$\ga$ 10 Myr (van den Ancker, private communication). 

\begin{figure}
\resizebox{\hsize}{!}{{\rotatebox{90}{\includegraphics{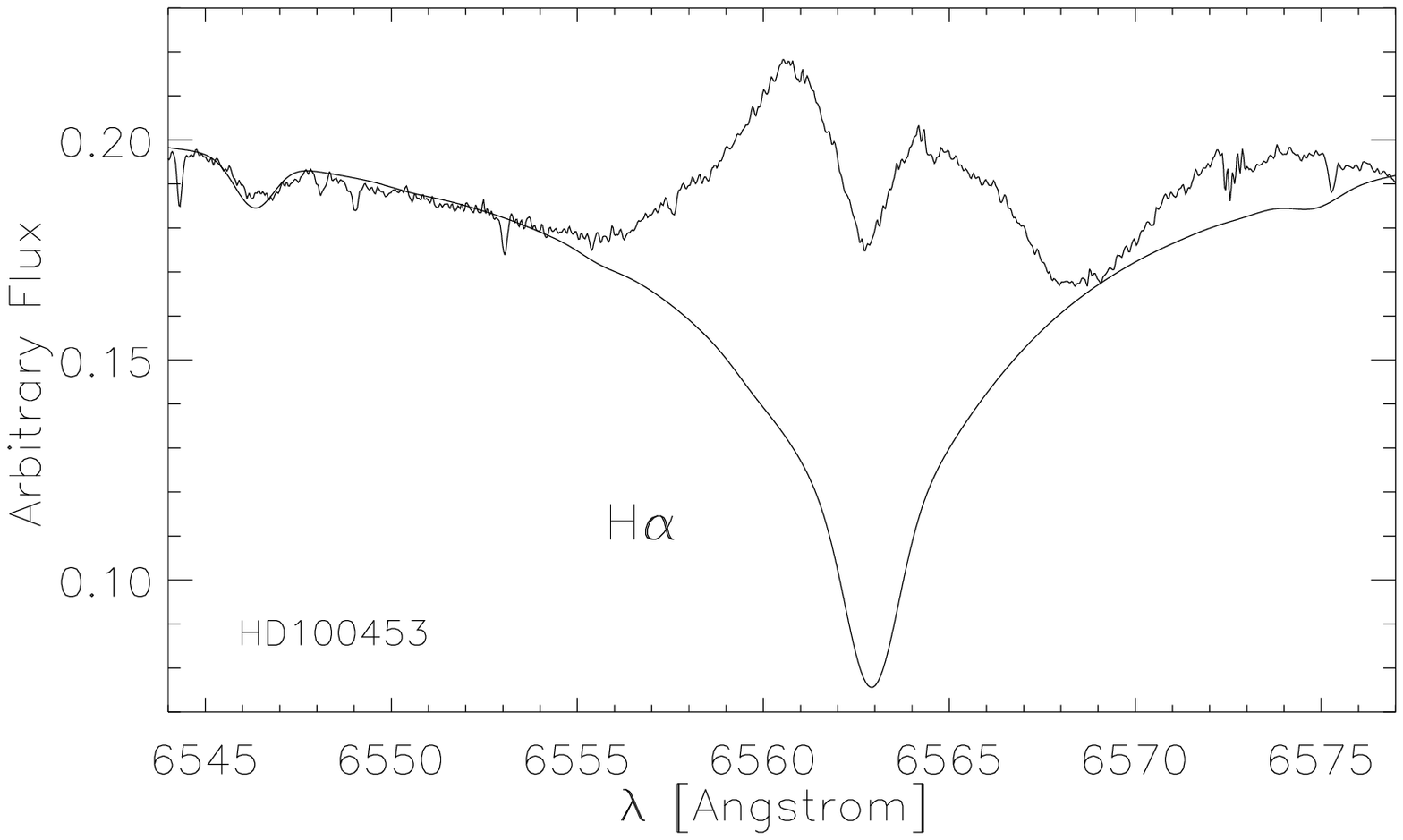}}}}
\resizebox{\hsize}{!}{{\rotatebox{90}{\includegraphics{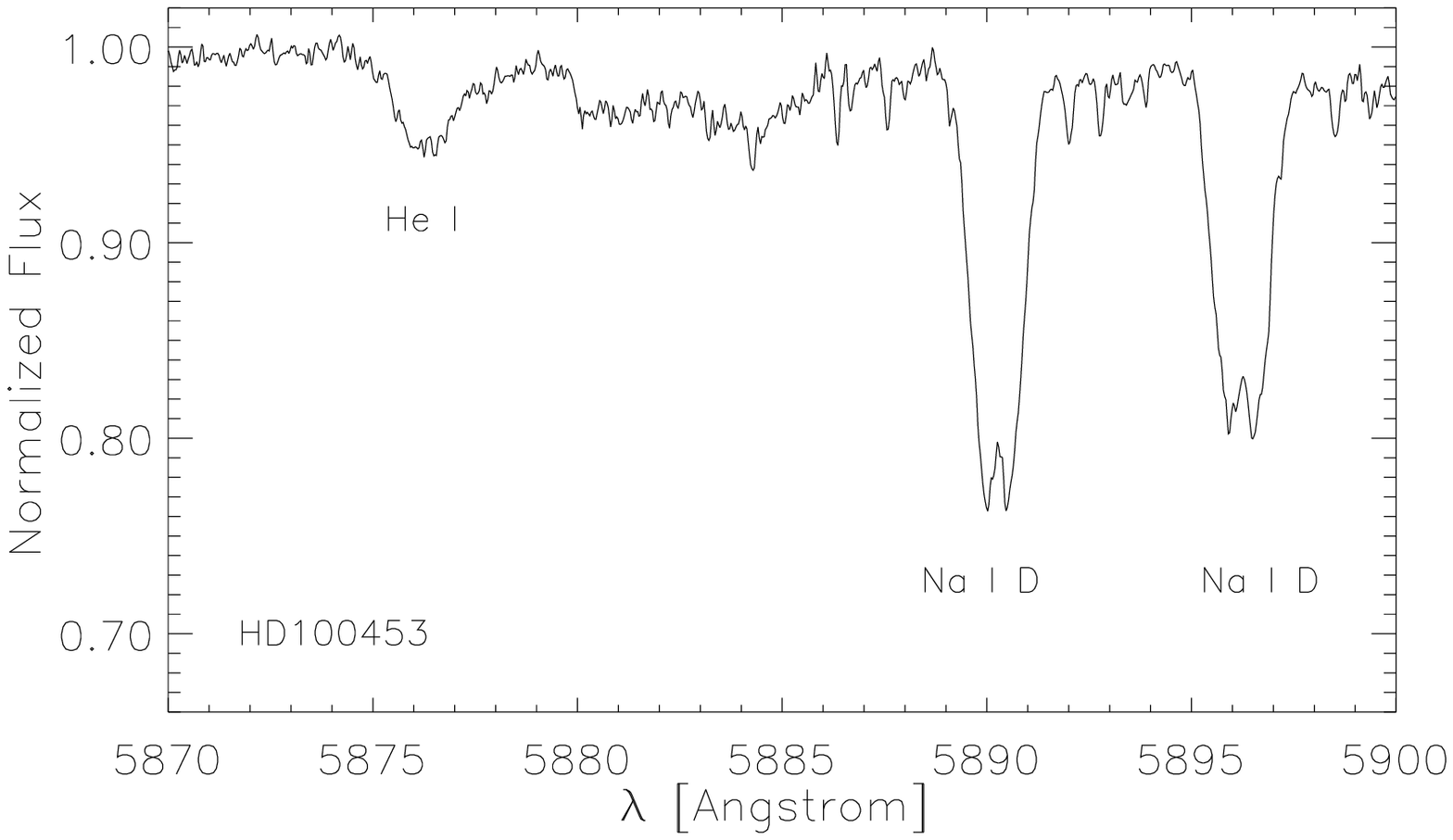}}}}
\vspace{-0.6cm}
\caption{Upper panel: The \halfa \ line in the spectrum of HD~100453, showing a 
double-peaked profile, and an additional redshifted emission feature. The 
absorption profile shows what is expected for an A9-type MS star. Lower panel: 
The spectrum of HD~100453 at $\lambda$ $\sim$ 5885 \AA. The Sodium lines 
are stellar, and show weak core emission in the centre, while the \HeI \ 
absorption line must be circumstellar, as its presence is not expected for an 
A9-type star.}
\label{spec}
\end{figure}

HD~100453 was observed in February 1999 with the echelle spectrograph Feros on 
the ESO 1.52m telescope at La Silla. We obtained a spectrum covering the 
wavelength range between 3700 and 9220 \AA. Earlier studies of HD~100453 had 
not yet established its HAEBE membership, but the spectrum we obtained in the 
\halfa \ region (Fig.~\ref{spec}, upper panel) readily confirms
its emission-line character. The \halfa \ profile is double
peaked, and it shows an additional redshifted emission component. About 50\% of 
the HAEBEs also show a double-peaked emission profile 
\citep{finkenzeller1984A&AS...55..109F}. In the lower panel of Fig.~\ref{spec},
we display the region around 5850 \AA; it is typical for a young star
surrounded by a disc: the \HeI \ line appears in absorption, which is
not expected for the atmosphere of an A9-type star. We suggest that it
has a CS origin, probably from a hot region close to the stellar
surface, as the width of the line is similar to the rotational
velocity of the star. \citet{dunkin1997MNRAS.290..165D} suggest that
the presence of excess \HeI \ emission or absorption is associated
with accretion, bringing the impacting material to a higher
temperature. \citet{dunkin1997MNRAS.290..165D} further noted that
HAEBE stars with a double-peaked \halfa \ profile show \HeI \
$\lambda$5876 \AA \ in absorption, unlike stars with a single-peaked
\halfa \ profile, which show the same line in emission; this is
probably an inclination effect \citep[see
also][]{meeus1998A&A...329..131M}. Furthermore, the Sodium lines are
mainly stellar, as appears from a comparison between the observed
spectrum and a synthetic spectrum of a star with similar parameters;
they show an additional - probably CS - weak core emission, which has
also been observed in other HAEBEs
\citep{dunkin1997MNRAS.290..165D}. We conclude that the optical
spectrum of HD~100453 is consistent with a HAEBE nature for this star.

\begin{figure}
\resizebox{\hsize}{!}{{\rotatebox{90}{\includegraphics{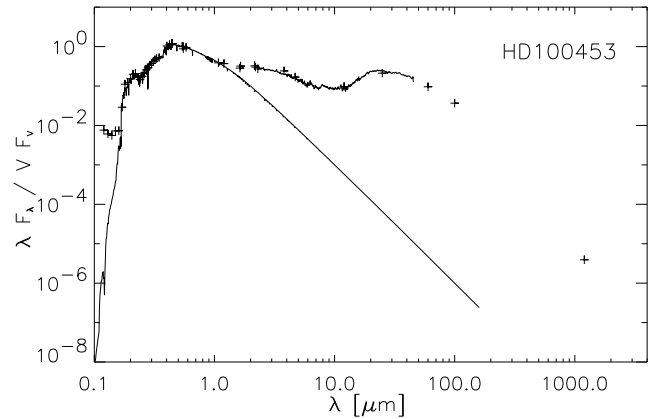}}}}
\vspace{-0.4cm}
\caption{The composed SED of HD~100453, showing a double-humped shape in the IR.
We overplotted the photometric data (crosses) with the SWS spectrum, as well as 
with a Kurucz model (straight curve), representing the stellar contribution to 
the SED.}
\label{sed}
\end{figure}

We constructed a SED for HD~100453, composed of optical, near-IR and IR data 
up to 100~\mic \ (see paper I for references), and a recent millimetre 
observation at 1.2 mm with SIMBA on the SEST (F$_{1.2\mathrm{mm}}$ = 30 $\pm$ 3 
mJy; S. Wolf, private communicaton); the SED displayed in Fig.~\ref{sed}. 
From this long-wavelength observation we derived a gas mass of M$_{\mathrm{g}}$ = 
3.0 $\times$ 10$^{-3}$ \Msun, following \citet{henning1998A&A...336..565H}.

HD~100453 was observed with ISO-SWS, in the full scan mode (mode 
AOT1); there is no LWS spectrum available. The SWS spectrum spans between 2 and 
45~\mic; it is relatively featureless, apart from the PAH features which are 
observed at 3.3, 6.2, 7.7 and 11.2~\mic. The spectrum is relatively flat in the 
short wavelength region, but then rises steeply longwards of 10~\mic. In our 
sample of 14 isolated HAEBEs \citep{meeus2001}, we defined two groups based
upon the shape of the SED in the IR: the geometry of group I sources was 
associated with an optically thin, flared region surrounding an optically thick, 
cold disc, for group II sources with a geometrically flat disc. These groups were 
further subdivided based upon the presence (a) or absence (b) of the 10~\mic \ 
silicate feature. HD~100453 is one of the four stars belonging to group Ib: the 
silicate emission feature is absent while the overall shape of the energy 
distribution is very much like that of other isolated HAEBEs which do show a 
silicate feature (group Ia). 
                
\section{Modelling}

We used an optically thin radiative transfer model, the code MODUST 
(Bouwman \& de Koter, in preparation) which was already succesfully used to 
reproduce the SWS spectra of several other isolated HAEBEs: AB Aur and HD~163296 
by \citet{bouwman2000A&A...360..213B}, HD~100546 and HD~142527 by 
\citet{malfait1998A&A...332L..25M,malfait1999A&A...345..181M}. These studies 
revealed several common characteristics of the CS dust around HAEBEs: first, the 
dust is distributed over two temperature regimes, a hot (T $\sim$ 1500-400 K) and 
a cold (T $\sim$ 300-20 K) one; most of the mass is contained in the cold 
component. Furthermore, the bulk of the mass ($\sim$ 70\%) consists of 
silicates. Other dust components such as carbonaceous material, iron oxide, 
metallic iron and water ice are present as well; some stars also show evidence 
for crystalline material (silicates and water ice). 
Both the density $\rho(r)$ and the size distribution $n(a)$ are assumed to have 
a power-law dependence of the radial distance $r$ and size $a$, respectively. 
For a thorough description of the dust model, we refer to 
\citet{bouwman2000A&A...360..213B}. 

\begin{figure}
\resizebox{\hsize}{!}{{\rotatebox{0}{\includegraphics{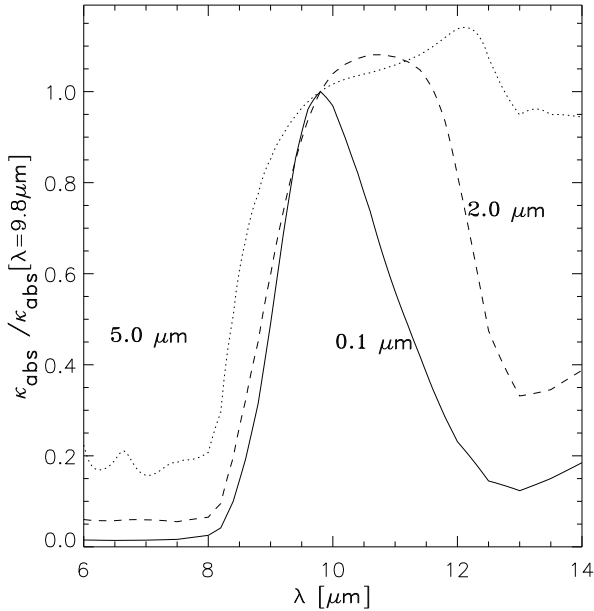}}}}
\caption{The emission properties of silicate dust grains, plotted against
wavelength $\lambda$. Be carefull, because in this plot the absorption 
coefficients are normalised. Shown are the coefficients of amorphous silicates 
for three grain sizes: 0.1, 2.0 and 5.0~\mic, with a normalisation factor of 
2377, 1540 and 568, respectively. 
The larger the grain size, the more the peak shifts to the right and the peak
strength weakens; when the grains 
reach a fairly large size ($\sim$ 5~\mic), the feature flattens and weakens
until it becomes impossible to distinguish it from the continuum.}
\label{kappa}
\end{figure}

In Fig.~\ref{kappa} we display the emission properties of silicate grains of
different sizes, to show the influence of the grain size upon the appearance of 
the 10 micron spectral feature. When the grains become larger, the peak shifts 
towards longer wavelengths, while the strength relative to the continuum
diminishes. From a certain grain size ($\sim$ 5~\mic), the feature is no longer 
distinguishable from the continuum. In the following, we will refer to particles
smaller than this size as {\em small particles}.

\subsection{Problems encountered while modelling}

A known problem when modelling the CS dust emission in the IR is that both 
the spatial distribution as the dust properties cannot be determined uniquely
by relying only upon IR spectral observations. This means that modelling
attempts will always yield a possible solution, but not a unique one. We 
therefore looked for constraints on dust properties and composition, rather than 
for the best fit. \citet{bouwman2000tesa.conf...63B} identified and discussed 
several degeneracies (e.g. different grain-size distributions and/or disc 
boundaries can reproduce the same spectrum, providing the dust has the same
average temperature) occuring when modelling CS dust; we refer to that study for 
a discussion.

\subsection{Fitting the spectrum of HD~100453}

An obvious way to hide the silicate feature is to make the silicate grains 
large or to restrict the amount of small, warm silicate grains. If, in such a
model, small carbon grains were still present, this would result in a dust model 
in which the emission of the carbon grains dominate. However, the coexistence of 
an average size distribution of carbon grains which is substantially smaller than 
that of the silicate grains is not physically likely. Indeed, carbon and silicate 
particles are expected to coagulate in the same amount, since they have similar 
surface properties and critical velocities for sticking to occur 
\citep{Chokshi1993}.
Therefore, we attempted to model the spectrum with similar size distributions for 
silicate and carbon grains, and in the cold component even for water ice. 
Furthermore, we searched for a solution with a continuous size distribution,
in which small silicates are not a priori excluded. 

In our modelling we ignored inclination effects. It is indeed possible to
remove the silicate feature assuming an edge-on orientation of the disc, but this 
would imply large optical and near-IR extinction. This is not observed, given the 
relatively small value of E[B-V] and the presence of a near-IR excess.

The lack of spectral bands in the spectrum of HD~100453 makes it difficult
to model, as the composition of the dust is virtually unconstrained. From
a comparison with other isolated HAEBEs, however, we know that the shape of the
IR energy distribution of HD100453 is typical for a HAEBE star (see paper I). 
Therefore, we took as a starting point the derived dust composition of AB Aur, 
which was also modelled with MODUST \citep{bouwman2000A&A...360..213B} and varied 
the parameters to fit the spectrum of HD~100453.
We studied the effect of the size and temperature distribution on the presence 
and strength of the silicate feature in two steps: (1) first we considered a 
similar -continuous- size distribution for silicate and carbon grains, and matched
that to the observed spectrum and determined the smallest average size still in 
agreement with the observed spectrum; (2) in a next step, we made an estimate of 
the maximum mass of small silicate grains that can be present, by adding 
single-sized small silicate grains to the previous fit. 

\subsubsection{Similar size distributions for silicate and carbon 
grains}\label{only}

It is clear that different species located within the same distance range can 
have a different temperature, because of their different absorption and emission 
properties. An important effect of adopting similar grain sizes while needing 
larger silicate grains to avoid the silicate feature is that the dust should be 
located relatively close to the star, so that it can become warm enough.  

We used MODUST with two dust shells as input, representing two different 
temperature regimes. The two distinct temperature regimes of the CS dust are a 
hot dust component with a mass averaged temperature of $\sim$ 360 K, and a cold 
dust component with a mass averaged temperature of $\sim$ 60 K. The hot region is 
located at a distance of $\sim$ 0.4, extending out to 3.4 AU; the cold regime 
starts at $\sim$ 8 AU, while for the outer radius a value of 96 AU is assumed. 
We stress that the distances adopted here and in the following parts for the hot 
and cold component are merely chosen to obtain a correct average dust temperature 
to fit the spectrum, and are probably not the actual dimensions of the shell due 
to the neglect of optical depth effects. E.g., the presence of a gap could be an 
artefact of our modelling and might not be necessary if a part of the disc is 
shielded from direct stellar radiation, for instance by an optically thick region.
This would also allow for two distinct temperature regimes. This geometry can be 
verified when using a 2-D model, allowing for a more complicated geometry and 
different optical thicknesses in different regions. 

\begin{table}
\caption{Model parameters of HD100453, using a similar size distribution for the 
carbon and silicate grains. Listed are the parameters defining the density 
and grain-size distribution, the chemical composition and the mass fraction 
M$_{\mathrm{frac}}$ of the individual dust species in both the hot as well as in 
the cold dust component. The uncertainties on the mass fractions of the different 
components are discussed below.}\vspace{1ex}
\label{para2}
\begin{center}
\begin{tabular}{r|rrrr}
\hline
\hline
\rule[3mm]{0mm}{1mm}
Component:         &\multicolumn{2}{c}{Hot}      &\multicolumn{2}{c}{Cold}\\[1ex]
\hline
\hline
\rule[3mm]{0mm}{1mm}
Radius (AU)        &\multicolumn{2}{r}{0.4--3.4} &\multicolumn{2}{r}{7.6--96}\\
$\rho \ (r)$       &\multicolumn{2}{r}{$\propto r^{-1.1}$}&\multicolumn{2}{r}{$\propto r^{-1.2}$}\\
$\rho_{0}$ [gr cm$^{-3}$]&\multicolumn{2}{r}{5.0 $\times$ 10$^{-15}$}&\multicolumn{2}{r}{8.0 $\times$ 10$^{-16}$}\\
M$_{\mathrm{dust}}$ (\Msun)&\multicolumn{2}{r}{2.9 $\times$ 10$^{-9}$}&\multicolumn{2}{r}{5.6 $\times$ 10$^{-6}$}\\
T range (K)              &\multicolumn{2}{r}{230--1500}&\multicolumn{2}{r}{20--200}\\
$n (a)$            &\multicolumn{2}{r}{$\propto a^{-3.5}$}&\multicolumn{2}{r}{$\propto a^{-3.5}$}\\[1ex]
\hline
\rule[3mm]{0mm}{1mm}
Dust species       &M$_{\mathrm{frac}}$&$a$ (\mic)&M$_{\mathrm{frac}}$&$a$ (\mic)\\[1ex]
\hline
\rule[3mm]{0mm}{1mm}
Olivine                  &0.76  &4--200              &0.76               &1.6--250\\
Carbon                   &0.16  &4--200              &0.13               &1.3--200\\
Water Ice                &--    &--                  &0.11               &1.6--200\\
Metallic Iron            &0.08  &0.01--2.5           &--                 &--\\
Iron Oxide               &0.003 &--                  &0.0001             &--\\[1ex]
\hline
\end{tabular}
\end{center}
\end{table}

\begin{figure}
\resizebox{\hsize}{!}{{\rotatebox{90}{\includegraphics{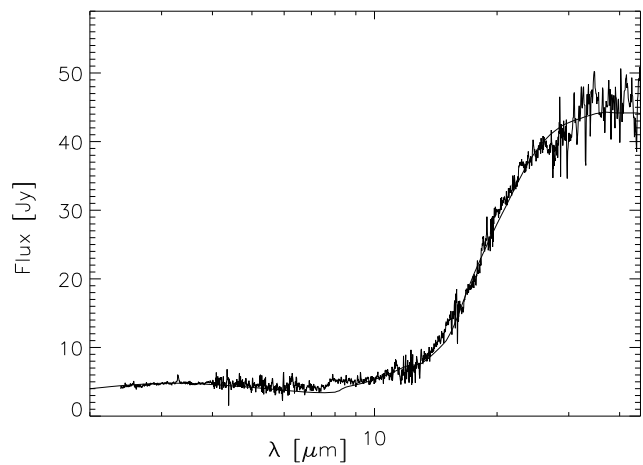}}}}
\caption{Modelling (straight curve) of the ISO-SWS spectrum of HD100453 (noisy 
curve). The parameters of this fit are listed in Table~\ref{para2}. 
Around 7--8~\mic, the model predicts too little emission; it is here that a 
broad PAH band is situated which we did not attempt to model.}
\label{fit}
\end{figure}

We derived a total dust mass in the hot component of 2.9 $\times$ 10$^{-9}$ 
\Msun \ and 5.6 $\times$ 10$^{-6}$ \Msun \ for the cold component. Most of the 
mass is thus in the cold dust component (the cold over hot dust mass ratio  
is $\sim 1.9 \times 10^{3}$). In Table~\ref{para2}, we list the parameters of the 
best model fit; in Fig.~\ref{fit} we show this fit, overlaid on the observed SWS 
spectrum. 

Because of the absence (apart from PAH bands) of spectral features in the spectrum
of HD~100453, constraints upon the chemical composition of the CS dust are 
difficult to make. Keeping the other parameters listed in Table~\ref{para2} as
a constant, we derived the following compositional limits for our {\em hot 
component}: amorphous olivine (70--76\%), amorphous carbon (8--16\%) and
metallic iron (8--12\%). Iron oxide might be present as well, but is not needed 
to obtain a satisfactory fit (0--10\%); its presence, however, is
likely because of the 
fast oxidation rates of metallic iron \citep{jones1990MNRAS.245..331J}. The 
presence of metallic iron is needed to explain the hottest near-IR region, as 
other species would not survive the high temperatures which are close to the dust 
destruction temperatures. For 
the {\em cold component}, next to amorphous silicates (70--90\%) and carbon 
(8--14\%), also iron oxide is likely to be present, though in very small amounts. 
Also water ice might be present in the cold component, but it 
is not required (0--20\%) to obtain a satisfactionary fit; we included it in 
analogy with modelling results of other isolated HAEBEs. The presence of metallic 
iron is highly unlikely in the cold component, given their relatively short 
oxidation timescale \citep{jones1990MNRAS.245..331J} at lower temperatures (T $ 
\leq$ 400 K). Crystalline silicates might be present, but will not reveal their 
presence in the spectrum when adopting the same size distribution as we assumed 
for the amorphous silicates; we therefore did not consider them in our simple 
approach.

As we discussed above, from a featureless spectrum, it is difficult if not 
impossible to determine the exact grain-size distribution. Furthermore, there
is a degeneracy in maximum grain size $a_{\rm{max}}$: when one moves the outer 
radius more in, $a_{\rm{max}}$ becomes larger, as larger grains at a smaller 
distance can produce a similar continuum as smaller grains at larger distance. 
For the purpose of this paper we consider the mean radius $<a>$,
defined as
\begin{equation}
\label{eq:1}
<a> = \frac{\int\limits_{a_{\rm min}}^{a_{\rm max}} a n(a) da}{\int\limits_{a_{\rm min}}^{a_{\rm max}} n(a) da}
\end{equation}
\cdrev{where $n(a)$ is the number of particles with size $a$.
We use this value because the maximum radius is only poorly
constrained due to the steep powerlaw of the size distribution
required to fit the spectrum.} From our modelling, some trends 
can be noted: the silicate grains around HD~100453 are substantially larger than 
what is found for the interstellar medium \citep[there $a$ is between 0.005 and
0.25~\mic, with a mean radius of $\sim$ 0.008~\mic;][]{mathis1977ApJ...217..425M};
the mean radius $<a>$ of the cold silicate grains is smaller (2.7~\mic) than 
that of the hot silicate grains (6.7~\mic). \cdrev{An important result is that the 
minimum radius of the smallest grains is larger in the hot component than in the 
cold component. This can be explained as follows: in the hot
component, small grain would cause a silicate emission feature,
therefore they cannot be present.  Grains in the cold component, however, are so
cold that no relevant contribution at 10 micron is produced by the
emission of these grains.
Therefore, small grains are allowed there without contradiction to the
observed spetrum.}

\subsubsection{Maximum mass of small, hot silicate particles}\label{secmax}

\begin{figure}
\resizebox{\hsize}{!}{{\rotatebox{90}{\includegraphics{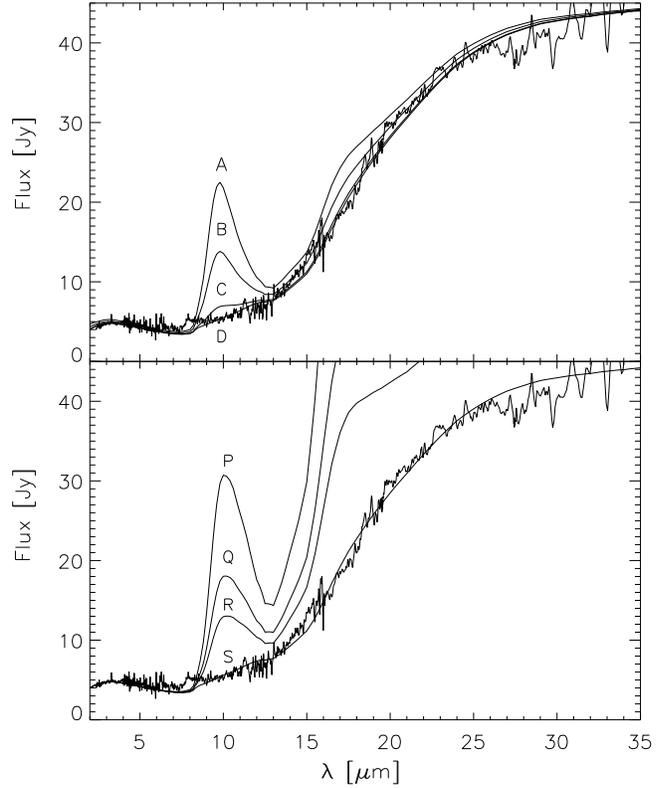}}}}
\caption{The effect of adding different amounts of mass of small silicate particles 
with $a$ = 0.1 to the previous fit. Upper panel: the effect in the hot component.
The added masses are, in \Msun, A: 6 $\times$ 10$^{-11}$, B: 3 $\times$ 10$^{-11}$, 
C: 6 $\times$ 10$^{-12}$ and D: 6 $\times$ 10$^{-13}$, being ad maximum 2\% of the 
hot dust mass. Lower panel: the effect in the cold component. The added masses are, 
in \Msun, P: 7 $\times$ 10$^{-8}$, Q: 3.5 $\times$ 10$^{-8}$, R: 2.1 $\times$
10$^{-8}$ and T: 7 $\times$ 10$^{-10}$, being ad maximum 1.2\% of the cold dust 
mass.}\label{figmaxmass}
\end{figure}

In the previous section, we showed that we could reproduce the spectrum of 
HD~100453 with a dust composition favouring larger ($<a> \ \sim $ 6.7~\mic) 
silicate grains; we will now show that small grains still can be present, albeit 
in a limited amount. Therefore, we started from our previous fit, and added 
grains of several characteristic sizes to this model: $a$ = 0.01, 0.02, 0.05, 
0.1, 0.2, 0.5, 1 and 2~\mic. For these grains, the same radial density dependence 
as the one used in the previous fit was adopted, with the same boundaries. For 
each selected silicate grain size, a certain amount of silicate was added to the 
fit, up to a value which is still in agreement with the observed absence of the 
spectral silicate feature; we always ran the model with just a unique size of 
small silicates. Herewith, we obtained for each selected size a maximum mass of 
the silicate grains which can be present.In Fig.~\ref{figmaxmass}, we show the 
effect of adding an increasing mass of silicate grains for $a$~=~0.1~\mic.

The results are shown in Table~\ref{maxmass}. An obvious trend is that, the 
larger $a$ becomes, the more mass can be added without causing a spectral feature.
In the cold component, the amount of mass that can be added is a factor of 1000 
larger than in the hot component, while the feature at $\sim$ 20~\mic \ is now
much more prominent. We also note the expected temperature drop of 
the silicates when they become larger. The maximum mass possibly residing in 
small, hot grains is $\sim$ 0.2 to 0.7\% of the hot dust mass, which is a 
negligible fraction of the total dust mass. 

\begin{table}[h]
\caption{Listed are for each selected silicate grain size, its maximum mass without
causing a silicate feature. We also give the average temperature of the grains, both 
in the hot and the cold component.}\label{maxmass}
\begin{center}
\begin{tabular}{r|cccc}
\hline
\hline
\rule[3mm]{0mm}{1mm}
Component: &\multicolumn{2}{c}{Hot}             &\multicolumn{2}{c}{Cold}\\[1ex]
\hline
\hline
\rule[3mm]{0mm}{1mm}
Size    &M$_{\mathrm{max}}$                   &$<$ T $>$    &M$_{\mathrm{max}}$            &$<$ T $>$ \\
(\mic)  &(10$^{-12}$ \Msun)          & (K)       &(10$^{-8}$ \Msun)    &(K)\\[1ex]
\hline
\rule[3mm]{0mm}{1mm}
0.01    &4.54                        &688        &0.63                 &103\\
0.02    &4.28                        &703        &0.76                 &105\\
0.05    &3.60                        &760        &0.76                 &110\\
0.1     &3.56                        &793        &0.88                 &113\\
0.2     &3.40                        &744        &1.89                 &109\\
0.5     &5.73                        &589        &2.52                 &97\\
1       &10.3                        &465        &5.05                 &86\\
2       &17.2                        &376        &7.57                 &74\\[1ex]
\hline
\end{tabular}
\end{center}
\end{table}

\section{Discussion}

\subsection{Comparison with the Herbig Ae star AB Aur}

In an earlier study, the spectrum of AB Aur has been modelled with MODUST
\citep{bouwman2000A&A...360..213B}. The overal shape of the infrared SED of AB 
Aur is quite similar to that of HD~100453, but unlike that star, it shows 
emission features of amorphous silicate, indicating the presence of small, warm 
silicate grains. To summarise, the hot component of AB Aur can be reproduced with 
a dust shell consisting for 71\% of small ($a$ = 0.01--5.0~\mic) silicate 
grains, with a mean radius of 0.02~\mic. The cold component also contains 
larger grains, up to a size of 126~\mic, resulting in a mean radius of 
0.09~\mic. We now compare the results of AB Aur with those of HD~100453, 
considering two questions:\\

\noindent
{\bf Question 1: Are there small silicates present around HD100453?}\\
We assumed a silicate grain size distribution similar to that of carbon, and did not 
set a lower grain size limit a priori. We derived that, in order to avoid the spectral
feature to become visible, the minimum mean radius in the hot component needs to be 
$\sim$ 6.7~\mic, with a lower grain size limit of 4~\mic. This mean radius is 335 
times larger than the one derived for the hot silicate grains in AB Aur; this means
that in this case, the {\em size} of silicate grains in HD~100453 should be much 
larger than in AB Aur.\\

\noindent
{\bf Question 2: How many small, hot silicates can be present in HD~100453?}\\
\cdrev{For AB Aur, the mass of the small, hot silicate grains is 1.9 $\times$ $10^{-9}$ 
\Msun, with a ratio of small, hot silicate mass over hot dust mass $\sim$ 71 $\times$ 
$10^{-2}$; this is much more than what is found for HD~100453.  In
this source we find that the ratio 
of the mass of small, hot silicates over the total hot dust mass to be
at most $\sim$ 0.2--0.6 $\times$ 
10$^{-2}$. We can conclude that the {\em mass} of small, hot silicate grains around 
HD~100453 is \cdrev{at least} a factor 100 to 500 times less than for AB Aur.}

\subsection{Lack of small silicate grains}

As shown above, the ISO-SWS spectrum of HD~100453 can be fitted with a model 
which lacks small grains. This lack of small grains may be caused by several 
mechanisms: (1) If the gas content of the disc is very low the small grains are 
removed by radiation pressure and/or Poynting-Robertson (P-R) drag; (2) Otherwise, 
the grains in the disc have experienced more coagulation than other systems with a 
similar energy distribution, leading to a strong depletion of small grains.  

Adopting the 
timescale for P-R drag given by \citet{backman1993prpl.conf.1253B}, we derive that for 
a system like HD~100453, it will take approximately 7 $\times \ 10^{6}$ yrs to 
remove grains up to a size of 10~\mic \ up to a distance of 10 AU from the star. 
Radiation pressure would act even more efficiently, and remove all grains wich are
smaller than a few microns in $10^{4}$ yrs, up to a distance of 1000 AU. Thus, the
age of HD~100453 (larger than 10 Myrs) suggests that, if there is only little gas 
left to mix with the dust particles, radiation pressure and P-R drag had enough 
time to remove small grains. However, the (low) luminosity of HD~100453 results 
in larger timescales for these processes than e.g. would be the case for AB~Aur 
(L $\sim$ 48 \Lsun; age $\sim $ 2--4 Myr), which still has small grains. Furthermore, 
there are HAEBE stars of a similar age as HD~100453 and with higher luminosities 
(e.g. HD100546, L $\sim$ 32 \Lsun) that also still show a lot of small, hot silicate 
grains. More important, the presence of PAHs 
around these stars require the presence of gas, which would make these dust grain 
removal processes inefficient. It is hence improbable that radiative small dust grain 
removal processes have taken place on a large scale in the disc around HD~100453.

This leaves coagulation as an alternative. However, it is hard to understand the 
presence of very small grains as PAHs in the immediate environment of HD~100453 
and also to require the silicate grains to have coagulated so efficiently, 
leaving a gap in the size distribution between the PAHs and the larger
grains.  \cdrev{A possible explanation would be that the PAHs are
  located further away from the star - due to the single-photon
  heating of these grains, the emission temperature of PAHs is
  independent of distance.  Therefore, the PAHs are not necessarily
  co-spatial with the hot silicate grains.
Even though coagulation may have removed small grains, there is no
evidence that it has produced particularly large grains in HD100453.
Recent millimetre observations (S. Wolf, private
communication) imply a spectral slope between the 100 micron and 
1.2~mm fluxes of $\alpha$ = -3.7; where $\lambda$ F$_{\lambda}$ $\propto 
\lambda^{\alpha}$.
Here we need to remark that including the 100 micron point in deriving the mm-slope 
-which was necessary because of a lack of more mm-observations- makes this slope on 
average 0.5 dex shallower than when calculated from mm-points alone, as we could 
deduce from our other observations of HAEBEs. This means that the derived slope is only
an upper limit and that the real mm-slope $\alpha$ probably lies between -4.4 
and -3.9. When we now compare this value with the other HAEBEs in our sample 
(Paper I), it is clear that most of these stars show a shallower 
long-wavelength slope ($\alpha$ between --4.28 and --2.65, with an average of 
-3.25), suggesting that in these objects larger grains are responsible for the 
long-wavelength emission than in HD~100453. This indicates that coagulation has not 
yet occured on a large scale in HD~100453, a fact which may be in
contradiction with the required efficient coagulation of small grains.}

\subsection{Lack of small, hot silicate grains}

In Sect.~\ref{secmax} we showed that the IR appearance of HD~100453 can be 
interpreted as a result of a lack of small silicate grains of a certain 
temperature range, which we estimated to be between about 1000 and 200 K. 
Interestingly, this possibility has received theoretical support 
through the new disc models by \citet{dullemond2001}. DDN show that the SED of 
HAEBE stars can be understood in terms of a disc which has an inner radius whose 
scale-height is a significant fraction of the disc inner radius. This puffed-up 
inner disc region results in substantial shadowing of the region outside of this 
inner rim. This is the region where one normally would find grains with temperatures
 optimal for producing the 10~\mic \ silicate bump. Indeed, a preliminary fit of 
HD~100453 with the model of DDN predict that, of all grains in the optically thin 
surface layer, there is a only 3.4 $\times \ 10^{-11}$ \Msun \ that is 
warmer than 200 K. This is a negligible amount in the required temperature range 
to cause the silicate emission feature, and a number that is consistent with our 
empirical limits. \cdrev{However, in order to efficiently suppress the 10~\mic \ 
silicate band, the DDN models require either an artificially large scale-height
of the inner disc region or a gap in the disc at the location where the silicate 
feature normally is produced (Dominik et al 2002, in preparation). It is currently 
not clear if and how such a high scale height can be achieved. We conclude that the 
shielding of part of the inner disc by a high inner rim is an interesting alternative 
explanation for the lack of 10 micron silicate emission in  
certain HAEBE stars. This idea can be tested by obtaining high spatial resolution 
near-IR images, e.g. with AMBER on the VLT interferometer.}

\section{Conclusions}

We searched for an explanation for the absence of the 10~\mic \ silicate feature 
in the spectrum of HD~100453 in terms of either a size or a temperature 
effect. We first showed, from a study of the optical spectra, that the star 
belongs to the group of the isolated HAEBEs. HD~100453 is member of group Ib, 
which contains stars with an increasing IR SED that lack the silicate 
emission feature (see paper I). Our conclusions can be 
summarised as follows:

\begin{enumerate}

\item{we confirmed the HAEBE nature of HD~100453 with the observation of the 
\halfa \ line in emission, as well as the presence of other typical CS HAEBE 
lines; furthermore, the star has an age of approximately 10 Myr and is located 
close to the ZAMS}

\item{we modelled the IR spectrum of HD~100453 with an optically thin radiative
transfer code, with the following properties:

 \begin{itemize}

 \item the dust emission stems from two different temperature regimes, with an 
       average temperature of $\sim$ 360 K and $\sim$ 60 K; the bulk of the 
       mass consists of amorphous silicates, but carbonaceous material and 
       iron(oxide) are present as well

 \item the absence of the silicate emission feature can be caused by the absence
 of {\em small} particles: a size distribution starting at 4~\mic, with an 
 average of 6.7~\mic \ does not reveal the silicate feature. The minimum average
 radius derived for the silicate particles is about 355 times larger than that 
 for AB Aur, suggesting that, in the disc of HD~100453, grain growth has taken 
 place on a much larger scale than in the disc of AB Aur. 

 \item we exclude dust removal processes such as radiation pressure and P-R drag 
  to be the main cause for the difference in grain size, given the larger 
  luminosity of AB Aur and, more important the presence of CS gas which makes 
  this processes inefficient 

 \item the presence of small (hot) particles of species other than 
  silicates (such as PAHs and metallic iron) is established. Furthermore, from
  a comparison of the slope at longer (IR-mm) wavelengths with other HAEBEs 
  showing the silicate feature, we deduce a lack of large grains in the CS disc 
  of HD~100453. \cdrev{These may be seen as arguments against the coagulation
    explanation for the absence of the silicate feature.}
  
\end{itemize}
}

\item{alternatively, the absence of the silicate emission 
feature can be caused by the absence of a large amount of small, hot silicates; colder 
(T $<$ 200 K), small silicates still can be present in a much larger amount. We 
obtained a maximal mass of small, hot silicates between 10$^{-12}$ and 10$^{-11}$ 
\Msun, which is 0.2 to 0.7 \% of the hot dust mass, and roughly a factor 10$^{-6}$ of 
the total dust mass. When comparing HD~100453 with AB Aur, the maximal mass residing 
in the small, hot grains is a factor hundred to five hundred smaller. The derived
maximal mass is in agreement with the theoretical predictions by DDN.
They propose shielding by a puffed-up inner region as an attractive explanation for 
the absence of hot, small silicates; and hence for the absence of the 10 micron 
silicate emission feature.}

\end{enumerate}

\noindent
High spatial resolution observations are needed to give constraints on the 
location of the grains emitting in the infrared; also more long-wavelength 
observations are necessary to study the coldest dust. Only then will we be able 
to clearly distinguish between both effects. 

\begin{acknowledgements}
We thank M. van den Ancker for providing fundamental parameters, T. 
Reyniers for help on some of the calculations, and S. Wolf, for observing HD~100453 
with SIMBA/SEST. Part of this research was funded by the EC-RTN on 
``The Formation and Evolution of Young Stellar Clusters'' (RTN--1999--00436, 
HPRN--CT--2000--00155) and by the DLR grant 50\,OR\,0004. CD, LBFMW and AdK 
acknowledge financial support from NWO Pioneer grant 6000-78-333. This research 
has made use of NASA's ADS Abstract Service. 

\end{acknowledgements}

\bibliographystyle{/z/gwen/ARTICLES/A_and_A/bibtex/aa} 
\bibliography{/z/gwen/ARTICLES/A_and_A/bibtex/bibliofile} 

\end{document}